\documentclass[twocolumn,aps,pre,reprint,groupedaddress]{revtex4}
\usepackage{hyperref}
\usepackage{amsmath, graphicx, xcolor}
\usepackage{subfigure}

\begin{document}


\title{Resource dependency and survivability in complex networks}


\author{Madhusudan Ingale}
\author{Snehal M. Shekatkar}
\email[]{snehal@inferred.co}
\affiliation{Department of Scientific Computing, Modeling and Simulation, Savitribai Phule Pune University, Pune 411007, India}



\begin{abstract}
Components in many real-world complex systems depend on each other for the resources required for survival, and may die of a shortage. These patterns of dependencies often take the form of a complex network whose structure potentially affects how the resources produced in the system are efficiently shared among its components, which in turn decides a network's survivability. Here we present a simple threshold model that provides insight into this relationship between the network structure and survivability. We show that, as a combined effect of local sharing and finite lifetime of resources, many components in a complex system may die of lack of resources even when sufficient amount is available in the system. We also obtain a surprising result that although the scale-free networks exhibit a significantly higher survivability compared to their homogeneous counterparts, a vertex in the later survives longer on average. Finally, we demonstrate that the system's survivability can be substantially improved by changing the way vertices distribute resources among the neighbours. Our work is a step towards understanding the relationship between intricate resource dependencies present in many real-world complex systems and their survivability. 
\end{abstract}


\maketitle

\section{Introduction}
Many real-world complex systems take the form of a network of individual components that interact with other components in a complex fashion. Some prominent examples are the network of metabolites in the living cell, the World Wide Web, transportation networks, and social systems \cite{newman2006structure, newman2018networks}. The interactions between the components in these systems can take various forms depending upon the system; in the metabolic network the interactions are chemical reactions, while in the social network these could be friendships, acquaintances etc. Another important type of interactions that exist in these networked systems is `dependency' by which we mean that a vertex might depend on other vertices for the resources required for its survival. In a social network the resources could be various types of supports (monetary, professional, emotional etc.), while in the transportation networks, these could be commodities, technologies etc. A vertex may depend on other vertices for a particular resource either because it is incapable of producing the resource on its own, or because the amount produced at the vertex is not sufficient. The shortage of several required resources at a vertex often leads to its death or non-functionality. This in turn affects its neighbours in the network since they stop receiving the resources from the vertex that dies. It is easy to see that such high dependency in the system may lead to a cascade resulting into the death or dysfunction of the whole system. Hence, it is important to study the factors that affect the distribution of the resources in the network, and to devise strategies to improve it. 

Interestingly, in the field of business management, the problem of resource dependence is well-explored and goes by the name of `Resource dependence theory' put forth in a seminal work by Pfeffer and Salancik \cite{pfeffer2003external}. This theory addresses the question of how various organizations depend on each other for various resources, and how that affects their behaviour \cite{hillman2009resource, davis2010resource}. However, somewhat surprisingly, the network angle of the theory has remained largely unexplored in spite of evidently being highly relevant for understanding the collective behaviour of the organizations. Here we try to fill this gap by formulating a complex network theory of resource dependence. Also, because of its general nature, the theory is actually applicable not only to business management but to all types of systems where resource dependencies are present. 

It is worth noting that studies of different types of dependencies have been extensively done in the field of interacting or interdependent networks \cite{buldyrev2010catastrophic, gao2012networks, brummitt2012suppressing, zhang2020asymmetric}. We stress that these are different types of dependencies than presented here. Those dependencies could be thought of as ones in which stochasticity is only in the structure of the network, and effect of resources is deterministic, like the interdependencies between electric power stations and vertices in the Internet communication network \cite{buldyrev2010catastrophic}. In this case, failure of the power station is assumed to lead to failures of Internet communication vertices with certainty. Vertex dependencies discussed here, however, are of very different type in which vertices share stochastically produced resources with each other, and die whenever resource amounts are less than certain threshold. Also, the problem of dependencies in a single network, arguably being easier to work with, has not been paid much attention to in the literature. The threshold-based model presented here is for the study of resource dependencies in a single network. We find that the model shows a rich behaviour depending upon the structure of the underlying network and the way the resources are distributed among neighbours. We first analytically find the expected amount of time the network survives when the vertices are not allowed to share surplus resources with the neighbours, and then show how this time drastically increases when sharing is allowed. 

The rest of the paper is organized as follows. In section \ref{model}, we present a simple model to study resource dependencies in an isolated network, and obtain analytical results for the case when sharing resources with neighbours is not allowed. Then in section \ref{with_sharing} we discuss the effect of degree distribution on the network survivability when sharing is allowed. In section \ref{nonuniform}, we discuss a strategy that increases the survival time of the network. Finally we discuss possible future directions, and conclude in section \ref{conclusion}. 
\section{\label{model}A simple model of resource dependencies}
Consider an undirected, unweighted simple graph $G$ with $n$ vertices and $m$ edges. Each edge in the network depicts a potential flow of resources between the vertices it connects. Each vertex in the network may need different resources $A, B, C, \cdots$ to survive, some or all of which it can produce on its own. If a vertex doesn't produce a particular resource that is necessary for survival, or produces it in insufficient amount, then to stay alive, that resource must be imported from the neighbours. In the present work, we restrict ourselves to only a single resource which each vertex is capable of producing, and so the only thing that matters for survival is whether the vertex has a sufficient amount at any given time. 

To be concrete, we assume that time is discrete. Let's consider an instance of time t at which, a vertex $i$ in the network could either be alive or dead. If $i$ is alive, it produces amount $X_i(t)$ of the resource. Here $X_i(t)$ is a random variable with probability distribution $p(x; \theta_i)$ where $\theta_i$ represents the set of parameters of the distribution. We also assume that the vertex $i$ needs minimum $R_i$ amount of the resource to survive at each time step. We define the surplus amount at vertex $i$ to be $S_i(t) = X_i(t)-R_i$ when $X_i(t) > R_i$ and $0$ otherwise. If $S_i(t) > 0$, it is distributed among the neighbours of $i$ that are alive at time $t$. The neighbours that are dead receive zero share from the surplus, and hence if all neighbours of $i$ are dead, the surplus is simply discarded. The surplus amount need not be distributed equally among the neighbours. In fact as we show in this paper, the way this excess amount is distributed among the neighbours is one of the critical factors deciding the survivability of the system. On the other hand, if $X_i(t) \leq R_i$, the vertex keeps the whole amount to itself. Thus, the total amount $X_i^{\text{tot}}(t)$ at time $t$ is the sum of the amount left at the vertex after distributing to the neighbours, and the total amount received from the neighbours. If $Y_{ij}(t)$ is the amount received by $i$ from neighbour $j$, then we have:

\begin{equation}
X_i^{\text{tot}}(t) = \begin{cases}
X_i(t) + \sum\limits_jA_{ij}Y_{ij}(t) \quad \text{if} \quad X_i(t) < R_i\\
 \\
R_i +  \sum\limits_jY_{ij}(t) \quad \quad\text{otherwise}
\end{cases}
\end{equation}
where $A_{ij}$ is the $(i,j)^{\text{th}}$ element of the adjacency matrix of network. Since here we are working with unweighted simple graphs, $A_{ij}\in \{0, 1\}$.
        \begin{figure}
            \centering
            \includegraphics[width=0.95\columnwidth, trim = 0 0 0 0, clip = true]{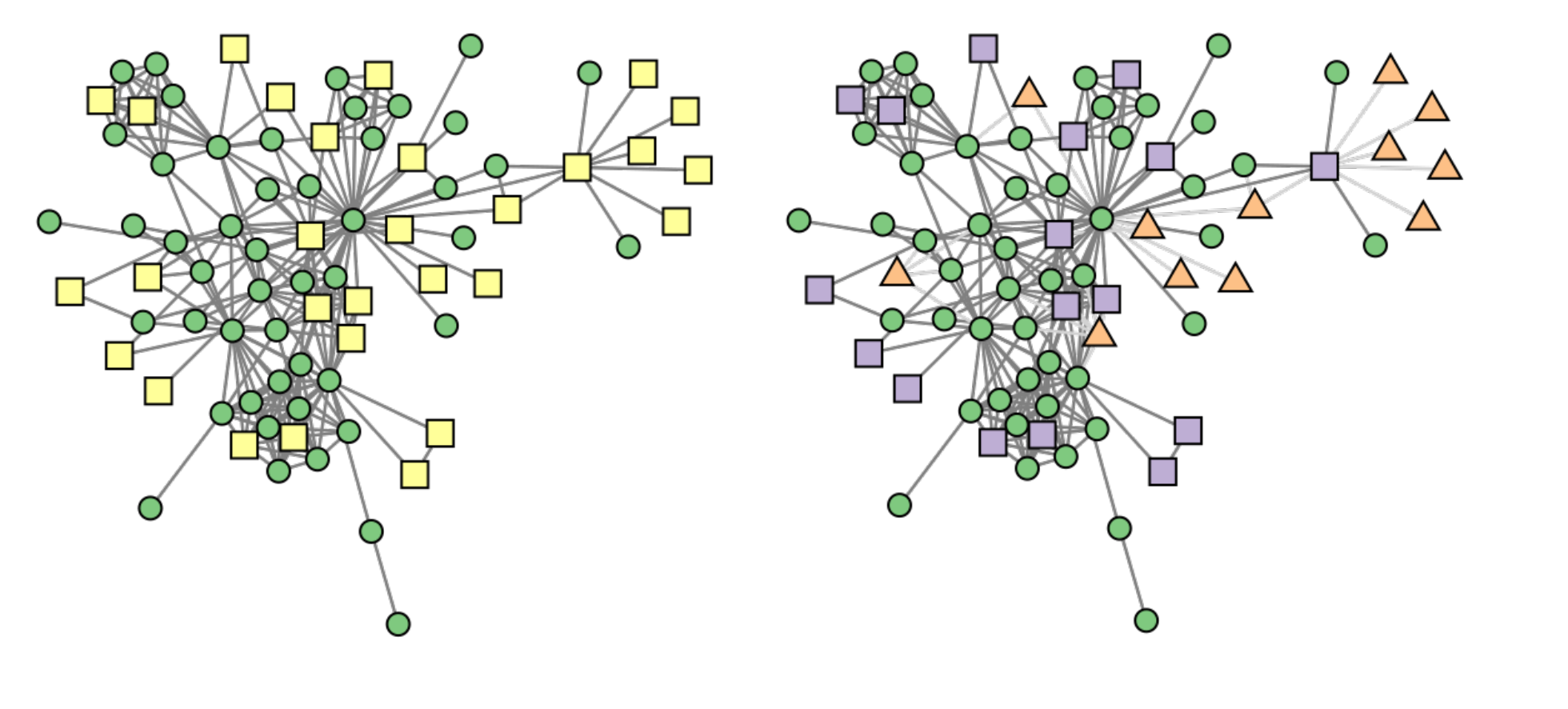}
            \caption{\label{resource_dynamics} Graphical illustration of the resource dependency model described in Sec.\ref{model} using the network of characters in \textit{Les Miserables} \cite{knuth1993stanford}. Left: each vertex produces the amount at time $t$. The vertices that produced the amount greater than the threshold $R_i$ are safe (green circles), whereas those who produced less than $R_i$ become vulnerable (yellow squares). Right: safe vertices distribute their surplus among the neighbours, and some of the vulnerable vertices receive enough amount to have total greater than $R_i$, and are saved (purple squares), while others die (orange triangles).}
        \end{figure}

If $X_i^{\text{tot}}(t) < R_i$, the vertex doesn't have sufficient amount to survive, and it immediately dies; else the vertex consumes all the available amount  at time $t$ although only $R_i$ is required. Below we make following assumptions about the resources explicit. 
    \begin{enumerate}
    \setlength\itemsep{1em}
       \item{First, we assume that the resource has a lifetime of $1$ time unit, and because of finite lifetime, a vertex cannot store it for future even when the amount at any time is greater than $R_i$. } 
       \item{Second, we assume that each vertex has only the topological information about its neighbours (for example, their degrees), and doesn't know how much amount is produced on them at any time. As we will see this has an important consequence when we try to devise strategies to improve the survivability of the network. } 
    \end{enumerate}

Fig.~\ref{resource_dynamics} graphically illustrates the model. When the vertices produce the resource, if the amount of produced at a vertex is greater than $R_i$, the vertex is in ``safe'' state indicated by green circles. Otherwise, the vertex is in ``vulnerable'' state shown by yellow squares. If a vulnerable vertex receives enough amount from the neighbours, it goes into ``saved'' state (purple squares), or dies otherwise (orange triangles).
To simplify the analysis, henceforth we set the survival threshold $R_i$ for all the vertices to the same value, $R$.
\subsection{Survivability analysis without sharing}
First, let us consider the case where the vertices in the network are not allowed to share resources with each other, so that $Y_{ij}(t) = 0$ at all $t$. Thus, a vertex would die at time $t$ whenever $X_i(t) < R$. The probability that the resource amount is greater than the threshold $R$ is

\begin{equation}
\label{phi_eqn}
\begin{aligned}
\phi_i(R, \theta_i) = \int\limits_{R}^{\infty}p(x; \theta_i)dx
\end{aligned}
\end{equation}

Note that this probability depends on the parameters $\theta_i$, and so different vertices have different probabilities of surviving at any given time depending on their capability of producing the resource. The probability that the vertex $i$ dies at exactly time $t$ is thus:

\begin{equation}
\label{psi_eqn}
\begin{aligned}
\psi_i(t) = \phi_i^{t-1}(1-\phi_i)
\end{aligned}
\end{equation}
This expression just says that for the vertex to die at exactly time $t$, it must survive at all prior times $1, 2, 3, \cdots, t-1$ which happens with probability $\phi_i^{t-1}$ and then it must die at time $t$ which happens with probability $1-\phi_i$. 
Thus, if the vertices $1, 2, \cdots, n$ die at times $T_1, T_2, \cdots, T_n$ respectively, the total time for network death is given by:

\begin{equation}
\begin{aligned}
T = \text{max}(T_1, T_2, \cdots, T_n)
\end{aligned}
\end{equation}
The probability mass function (PMF) of $T$ can be calculated as follows. The probability that a vertex $i$ dies at time $T_i$ or less is given by the cumulative distribution:
\begin{equation}
\begin{aligned}
F_i(T_i) = \sum\limits_{t=1}^{T_i}\psi_i(t) = (1-\phi_i)\sum\limits_{t=1}^{T_i}\phi_i^{t-1} = 1-\phi_i^{T_i}
\end{aligned}
\end{equation}
Since $T_i \leq T$ for all $i$, the cumulative distribution of $T$ is given by:
\begin{equation}
\begin{aligned}
F(T) = \prod\limits_{i=1}^{n}F_i(T) = \prod\limits_{i=1}^{n}(1-\phi_i^{T})
\end{aligned}
\end{equation}
Thus, the probability that the network dies at exactly time $T$ is:
\begin{equation}
\label{isolated_final}
\begin{aligned}
f(T) = \prod\limits_{i=1}^{n}(1-\phi_i^{T})-\prod\limits_{i=1}^{n}(1-\phi_i^{T-1})
\end{aligned}
\end{equation}
In the special case when all the vertices are equally capable of producing the resources, $\phi_i$ is same (say $\phi$) for all the vertices, and the above expression simplifies to:
\begin{equation}
\label{fT}
f(T) = (1-\phi^{T})^n-(1-\phi^{T-1})^n
\end{equation}

In this paper we will focus only on the simple case where the parameters $\theta_i$ of the distribution $p(x; \theta_i)$ have the same values, $\theta_i = \theta$ for each vertex although this is not necessary in general. We use exponential distribution for the production of resources so that $X_i \sim \text{Exp}(\beta)$, with PDF given by:

\begin{equation}
p(x, \beta) =
\begin{cases}                                                              
    \frac{1}{\beta} e^{-x/\beta} & x \geq 0 \\                             
    0 & x < 0                                                              
\end{cases}
\label{expdist}                                                            
\end{equation}

Since the average of the distribution is $\beta$ for the Exponential distribution, higher the value of the parameter $\beta$, higher the rate of resource production. Using the form of the exponential distribution in Eq(\ref{phi_eqn}), it is easy to verify that in this case $\phi = e^{-R/\beta}$, and hence from Eq(\ref{fT}) we have:

\begin{equation}
\label{fT_exp}
f(T) = (1-e^{-RT/\beta})^n- (1-e^{-R(T-1)/\beta})^n
\end{equation}

\begin{figure}
	\includegraphics[width=0.95\columnwidth]{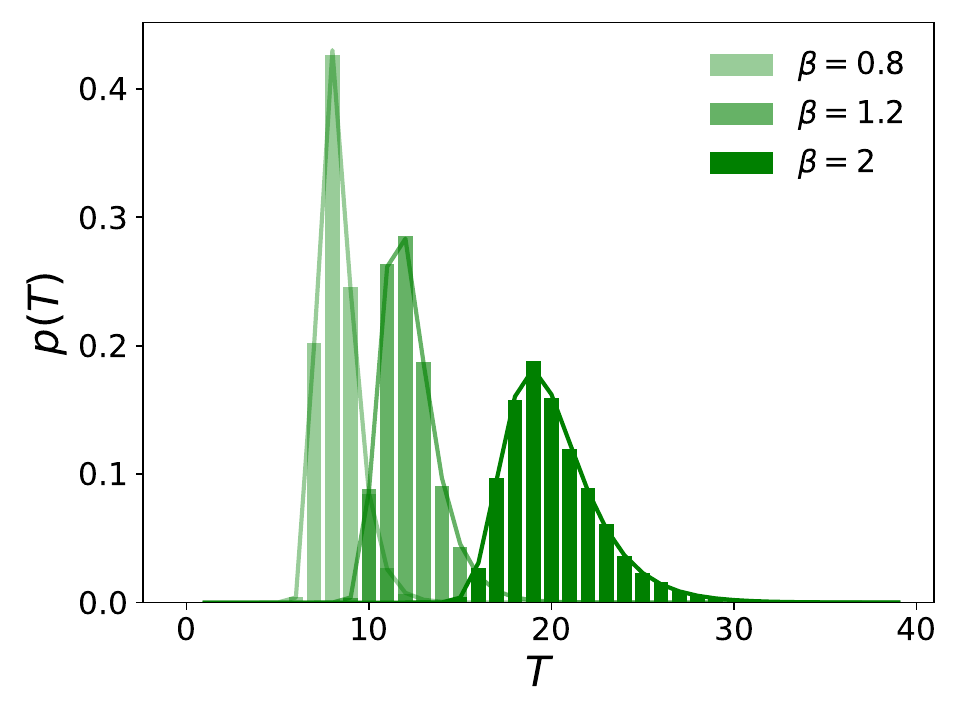}
	\caption{\label{no_sharing}The distribution of survival times for different values of $\beta$ with $R = 1$ and network size $n = 10^4$ when sharing resources is not allowed. The continuous lines are obtained analytically using Eq(\ref{fT}) while the vertical bars are obtained by simulation by averaging over $10^4$ realizations.}
\end{figure}

The distribution of the survival time for different values of $\beta$ obtained using this equation is shown in Fig.~\ref{no_sharing}. The bar heights in the plot show the numerical estimates, whereas the continuous curve is obtained analytically. At this point we also note that from Eq(\ref{fT}), $f(T)$ depends only on the ratio of the production capacity $\beta$ and the survival threshold $R$, and not on their absolute values. Physically this is obvious since if the threshold is large, and also the producing capacity is large, the chances of survival should remain unaffected. For this reason, in the remaining paper we fix $R = 1$ and $\beta = 2$. Here we also note that since $\beta > R$, the average amount per vertex produced in the network is greater than the amount required for survival. 
        
\section{\label{with_sharing} Survivability analysis with sharing of resources}
When the vertices with surplus $S_i(t) > 0$ are not allowed to transfer their surplus to other vertices (i.e. when $Y_{ij} = 0$), these surplus amounts are essentially wasted. This is because the corresponding vertices don't contribute to the survivability of vertices for which $X_i < R$. Thus, when we allow the sharing of resources (i.e. $Y_{ij} \neq 0$), the vertex $i$ dies only if $X_i^{\text{tot}} < R$, and hence we expect the average survival time $\langle T\rangle$ of the network to increase. In this section, we consider the case in which whenever the surplus $S_i(t) > 0$, it is distributed equally among the neighbours. The non-uniform distribution of the surplus is discussed in Sec \ref{nonuniform}. In the following analysis, we again use $X_i \sim \text{Exp}(\beta)$ as above. 

In this paper, we are primarily interested in the effect of the degree distribution on the network survivability. Therefore, as a substrate network, we use the configuration model which is a random graph model with a given degree sequence. In particular, we want to study how the survivability is affected by homogeneity or heterogeneity of the degree distribution of the network. To this end, we first construct the configuration model with the degree sequence drawn from the  Poisson distribution which is a homogeneous distribution. This is same as the famous Erd{\H o}s-R{\' e}nyi model when the network size is large. Similarly, to model the degree heterogeneity, we consider the configuration model with degree sequence drawn from a power-law distribution (henceforth called the \textit{power-law configuration model} or the \textit{scale-free graph}). 

Some discussion on the way the graphs are generated from the configuration model is in order. Traditionally, the configuration model is defined in terms of the ensemble in which every \textit{matching} of the half-edges is equally likely \cite{newman2018networks}. However, this implies that even the matchings that generate graphs with self-loops and multi-edges should be accepted. In our resource dependency model, inclusion of such graphs presents some problems in judging the dependence of graph survival on its structure. For example, a self-loop at vertex $i$ causes giving out less amount to the neighbours because now $i$ itself is one of its neighbours, and receives some amount from the surplus. Similarly a multi-edge causes extra bias in the distribution of surplus since a neighbour with two connections receives twice the amount than other neighbours when the surplus is shared equally among the neighbours. It is well-known that the density of the self-loops and multi-edges goes to zero in the limit $n\to \infty$ provided that the second moment of the degree distribution $\langle k^2\rangle$ is finite. Here apart from the Poisson random graph, we are also interested in the networks with power-law degree distribution, and since $\langle k^2\rangle$ diverges for the power-law case, asymptotically the density of self-loops and multi-edges does not go to zero. Thus, directly drawing graphs from the ensemble with uniform distribution over matchings is not a good idea for our purpose. For this reason, we resort to sampling from the ensemble of all simple graphs (i.e. graphs without multi-edges and self-loops) with a given degree distribution. While doing so, we must make sure that every such graph is drawn with uniform probability. This could be done using well-known Markov Chain Monte Carlo (MCMC) sampling procedure involving \textit{double edge swaps} \cite{fosdick2018configuring}. We use the implementation from \textit{graph-tool} to do this \cite{peixoto_graph-tool_2014}. 


As mentioned above, in this section we will consider the case in which whenever the surplus $S_i(t) > 0$, it is distributed equally among the neighbours that are alive at time $t$. First, as a model of networks with heterogeneous degree distribution, we construct the configuration model by drawing a degree sequence from a power-law degree distribution $p_k \sim k^{-\alpha}$ whenever $k \geq k_{\text{min}}$ and $0$ otherwise. This is the power-law configuration model or scale-free graph as stated before. In this paper, we fix $k_{\text{min}} = 2$ everywhere, and also restrict values of $\alpha$ between $2$ and $3$. The reason for this is two-fold : first, for most real-world scale-free networks, $\alpha$ values are seen to lie between $2$ and $3$, and second, the giant component surely exists in this range \cite{newman2018networks}. Then to compare the effect of such skewed distribution with a homogeneous distribution, we build the configuration model graph using the Poisson distribution with the average equal to the average of the corresponding power-law degree distribution. In the limit $n\to\infty$, this model is equivalent to the Erd{\H o}s-R{\'e}nyi graph. 

        \begin{figure}
            \centering
            \includegraphics[width=0.95\columnwidth, trim = 0 0 0 0, clip = true]{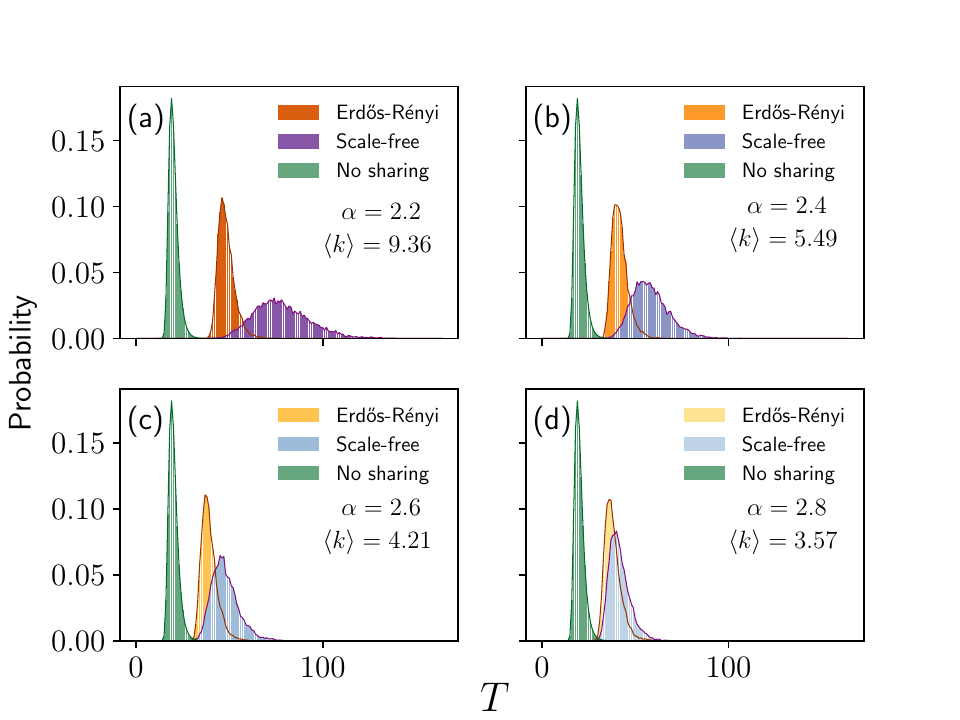}
            \caption{\label{T_dist}Probability distributions for the survival time for the scale-free (SF) and the Erd{\H o}s-R{\'e}nyi (ER) networks with the same average degree for different values of $\alpha$. The distribution for the `no-sharing' case is also shown for comparison. Smaller values of $\alpha$ correspond to higher-skewness for scale-free graph and to higher-density for the Erd{\H o}s-R{\'e}nyi graph. Sharing can be seen to drastically increase the survival time. Moreover, as the skewness and the density of the SF graph decreases, its survivability starts falling towards that of the ER graph. All the cases correspond to the network size $n=10^4$, $\beta=2$ and $10^4$ random realizations.}
        \end{figure}

\begin{figure}
            \centering
            \includegraphics[width=0.95\columnwidth]{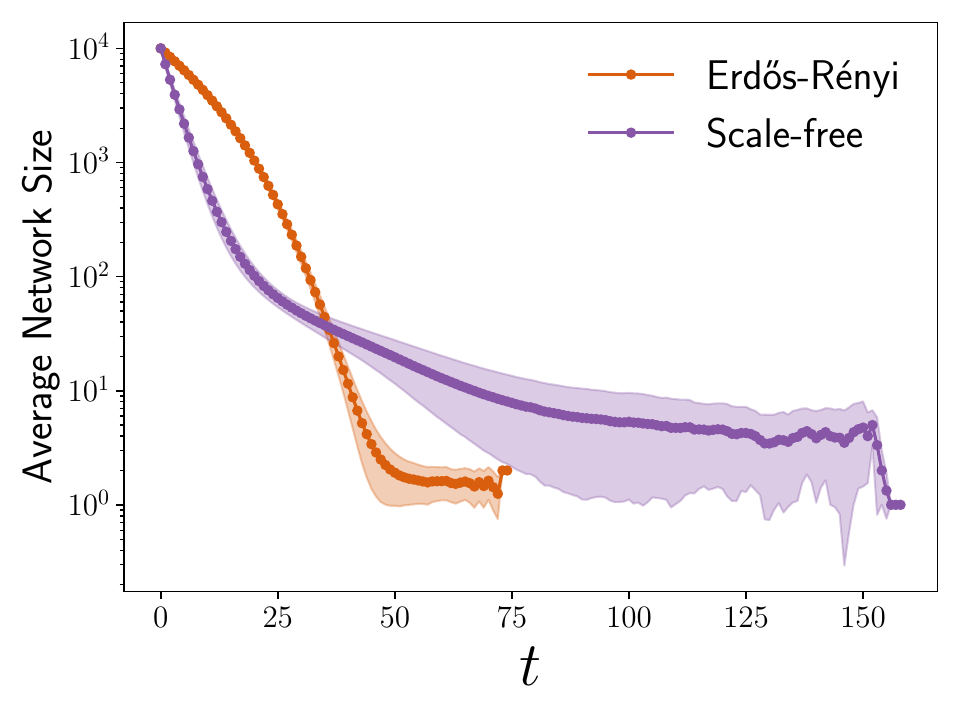}
            \caption{\label{tot_vert_vs_time} The semi-log plot of the average network size as a function of time for Erd{\H o}s-R{\' e}nyi and scale-free networks with $\beta = 2$ and $R = 1$. The scaling index of the SF graph is $\alpha = 2.2$, and the ER graph is constructed so as to have the same average degree as that of the SF graph ($\langle k\rangle = 9.36$). The results are averaged over $10^4$ random realizations for the network size $ n = 10^4$ each, and the bands around the curves are the data variations in terms of standard deviations.}
        \end{figure}

\begin{figure}
            \centering
            \includegraphics[width=0.95\columnwidth]{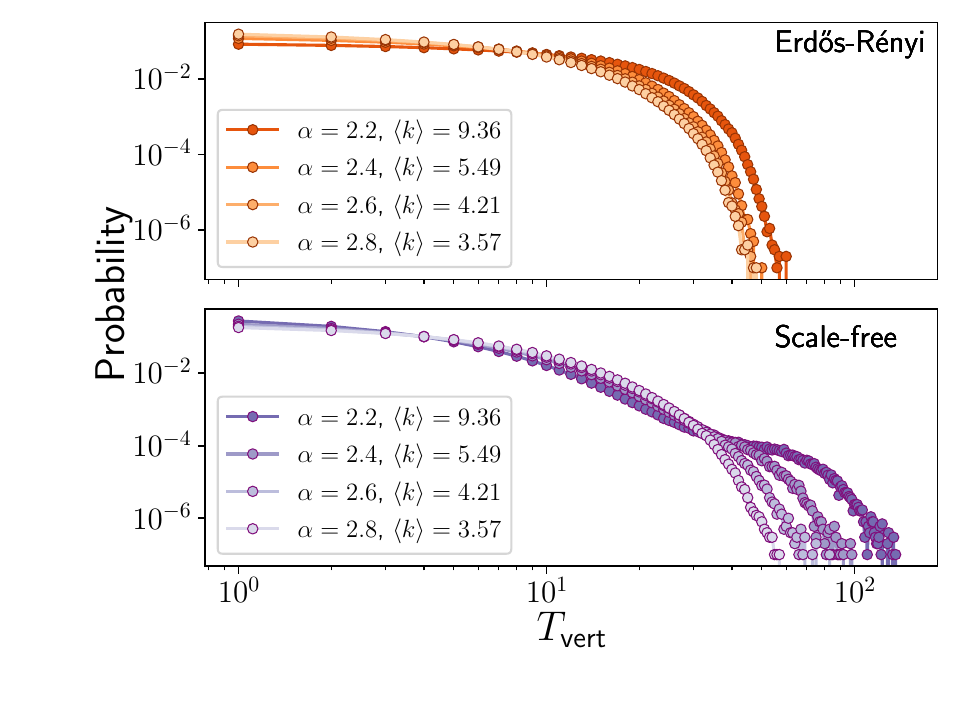}
            \caption{\label{time_dists} Distributions of vertex survival times for the ER and SF networks on log-log plot with $\beta = 2$. Notice the long-tail for the scale-free case that shows that some vertices in the network survive for substantially longer time than the average. The results are averaged over $10^4$ random realizations each of size $n = 10^4$ vertices.}
        \end{figure}

The distributions of the network survival times for the two cases are shown in Fig.~\ref{T_dist} which also shows the distribution for the no-sharing case with the same value of $\beta$. Clearly, the Scale-free graph survives for much larger times on average even when the two types of graphs have the same average degree and the same capacity to produce the resources (i.e. same $\beta$). To understand this phenomenon, we must study how the structures of the two types of graphs change with time. In Fig.~\ref{tot_vert_vs_time} we show the average network size as a function of time for the ER and SF topologies for $\alpha = 2.2$. We immediately see that initially most vertices in the SF graph die quickly. Comparatively, the decrease in the ER graph size is much slower. After a point, most vertices in the SF graph are deleted but a small fraction survives for long time, and this increases the survival time of the network. This is further corroborated by the plots shown in Fig.~\ref{time_dists} which show that for the SF network the distribution of survival times for individual vertices are themselves skewed. It seems reasonable that high-degree vertices are the ones that live longer, and we now present an explicit argument to support this. 

The total amount $Y_i(t)$ received by vertex $i$ from all its neighbours at time $t$ is: 
\begin{equation}
\begin{aligned}
Y_i(t) = \sum\limits_j A_{ij}(t)\frac{(X_j(t)-R)_+}{k_j(t)}
\end{aligned}
\end{equation}
where $(x)_+$ denotes the positive part of $x$ also known as Macaulay bracket. 
If we could derive the probability distribution of the surplus $Y_i$ from the above equation and the probability distribution of $X(t)$, it would be possible to derive the expression for the probability distribution for the survival time of the network. Unfortunately since $A_{ij}$ is time-dependent, this is difficult. 

Nevertheless, it is easy to see that the summation term in the equation above would have overall increasing trend since if we replace the neighbours' degrees $k_j$ by their average $\langle k\rangle_{\text{nbr}}$ and $X_j$ by their average $\langle X\rangle$, we get the following approximate expression : 
\begin{equation}
\begin{aligned}
Y_i(t) \approx k_i(t)\frac{\langle X\rangle-R}{\langle k(t)\rangle_{\textrm{nbr}}}
\end{aligned}
\end{equation}

\begin{figure}
            \centering
            \begin{center}
            \includegraphics[width=0.95\columnwidth, trim = 0 25 0 0, clip = true]{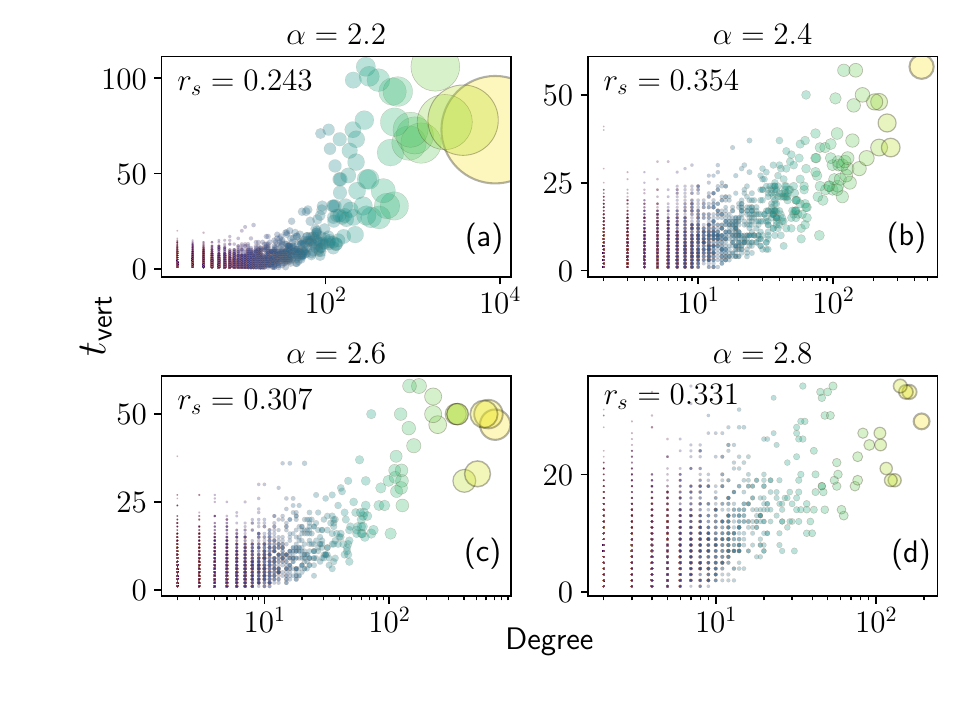}
            \end{center}
            \caption{\label{scatter_deg_time}Scatter-plots of the vertex survival times and their original degree values for the scale-free network of size $n = 10^4$ for different scaling indices $\alpha$. Results of only one realization are shown for the sake of better visualization, and larger circle sizes correspond to larger initial degree values in the network. The Spearman's rank correlation coefficient $r_s$ is seen to be positive showing the positive correlation as discussed in the text.}
        \end{figure}

Which is simply proportional to the degree. Thus, if a vertex has high degree initially, it receives higher amount from the neighbours on average, and has better chance of surviving longer. This results into skewed nature of the distribution of vertex times for the scale-free network since the degree distribution is itself skewed. To verify this, we look at the scatter-plot of the time $T_{\text{vert}}$ for which a vertex survives and its original degree as shown in Fig.~\ref{scatter_deg_time}. The positive correlation between the two is evident from the plots. However, we can't really expect the relationship to be completely linear since the actual time is decided not just by the original degree, but also by how it decreases in time. In the figure the trend looks roughly linear because the $x$-scale in the plots is logarithmic. For this reason, Pearson's correlation coefficient is a poor quantifier of the correlation between the two. Hence to quantify it, we use Spearman's rank correlation coefficient $r_s$ which only depends on the monotonicity of the data \cite{spearman1987proof}.

Thus, having more neighbours is always beneficial to a vertex since it then gets surplus amounts from many vertices which in turn increases its chances of surviving for a long time. This also means that if a vertex has low degree, it is extremely unlikely that it would survive for a long time. Thus, in the dynamics of resource dependencies, all the low-degree vertices in a scale-free network quickly die. This explains why initially the size of the SF graph decreases sharply compared to the ER graph as shown in Fig.~\ref{tot_vert_vs_time}.

\subsection{Average lifetime of a vertex }
\begin{figure}
            \centering
            \includegraphics[width=0.95\columnwidth, trim = 35 0 20 20, clip = true]{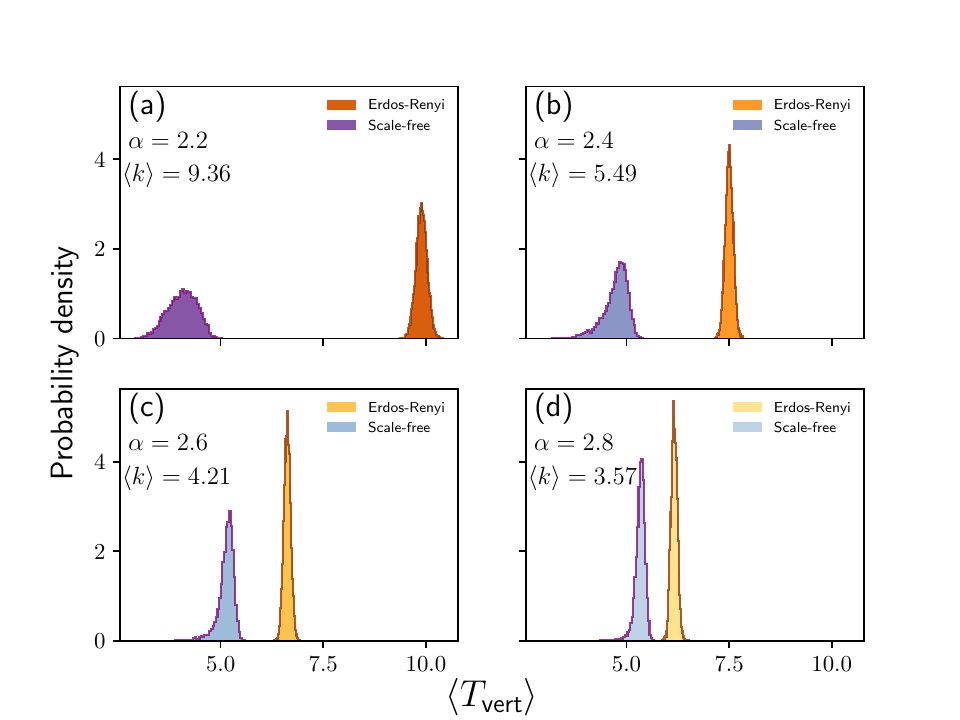}
            \caption{\label{ave_vert_t_unbiased} Histograms of average lifetime $\langle T_{\text{vert}}\rangle$ of a vertex in the network for Erd{\H o}s-R{\'e}nyi and Scale-free networks for different values of $\alpha$ obtained from $10^4$ random realizations each. Contrasting this with Fig.~\ref{T_dist}, we see that when skewness is high (low $\alpha$), although the scale-free network on an average survives longer than Erd{\H o}s-R{\'e}nyi network, a vertex in the later has a better chance of surviving longer. This is because the high survival time of SF networks is due to few high-degree vertices which survive much longer than the average. As the skewness decreases, the two types of graphs lead to similar values of $\langle T_{\text{vert}}\rangle$.}
        \end{figure}

The discussion so far may give the impression that Scale-free topology is substantially better in terms of surviving compared to homogeneous topology. However, we have also seen that the large survival time of the SF networks is due to a few high-degree vertices. This means that only a small part of the system actually survives long and the rest of the system dies rather quickly. From this point of view, it would be more relevant to compare the average amount of time for which a vertex in the ER and SF graph survives instead of the times for which the graphs survive. In analogy with Eq(\ref{psi_eqn}), let us define $\psi_i^{\text{share}}(t)$ to be the probability that a randomly chosen vertex in the graph survives up to time $t$ when sharing is allowed. We have already seen the distribution $\psi_i^{\text{share}}(t)$ in Fig.~\ref{time_dists}. The mean survival time of a vertex in the graph is given by:

\begin{equation}
\begin{aligned}
\langle T_{\text{vert}}\rangle = \sum\limits_{T = 1}^{\infty} T \psi_i^{\text{share}}(T)
\end{aligned}
\end{equation}
The distributions of this average time are shown in Fig.~\ref{ave_vert_t_unbiased} for the two types of graphs obtained using $10^4$ random realizations. This presents a drastically different picture than what we have been discussing so far, implying that the homogeneous ER graph should be treated as a much more survivable than the SF graph since a random vertex in ER network survives much longer on average. Since in the real-world systems we would like to make sure that a large part of the network can keep functioning, this result means that it is better to use homogeneous topology instead of heterogeneous topology.

\section{\label{nonuniform} Increasing network survivability using biased distribution of surplus resources}
As we discussed above, the high-degree vertices have higher chance of surviving owing to having many neighbours, and low degree vertices are extremely vulnerable. Now we ask an interesting question: is it possible to increase the survivability of the low-degree vertices while keeping the high-degree vertices safe? If this could be achieved, it would be possible to increase the survival time of the network without increasing the production of resources on the vertices. The main observation that lets us answer this question affirmatively is that high degree vertices in fact receive much more from the neighbours than the required amount $R$. Moreover, since the resource has a lifetime of only $1$ unit, the excess amounts cannot be saved for future, and are essentially wasted. If we could take some of those excess amounts and transfer those to the vertices in need of resource, we could make low-degree vertices survive longer. This could be achieved by distributing the surplus $S$ among the neighbours in a biased fashion so that the high-degree neighbours get less share of $S$ while the low-degree ones get higher share. 

\begin{figure}
            \centering
            \includegraphics[width=0.95\columnwidth]{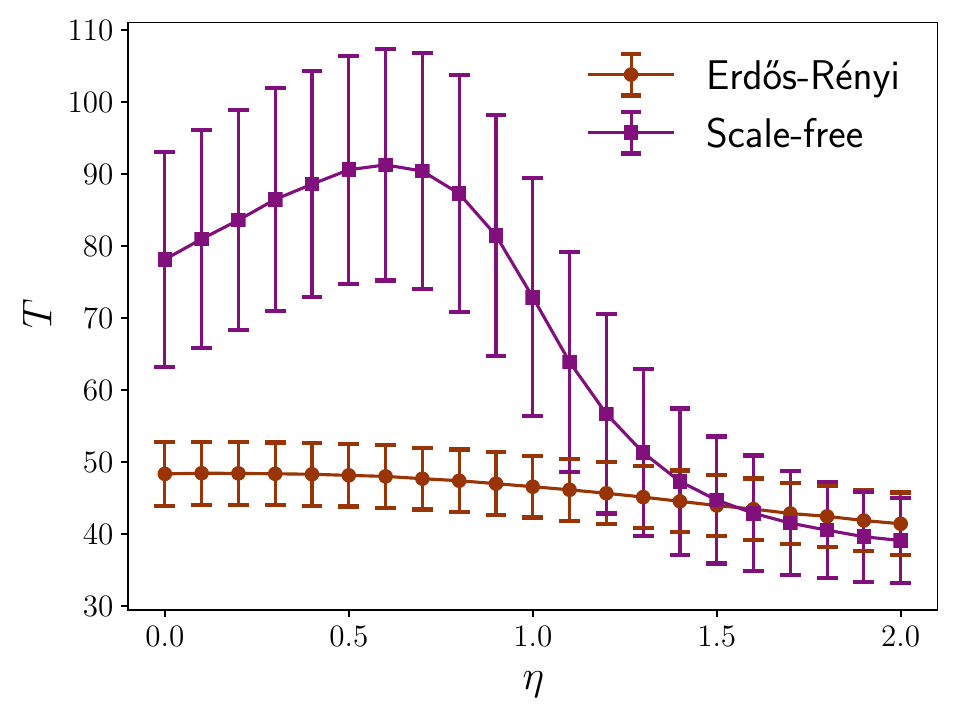}
            \caption{\label{t_vs_eta_beta2} The variation of the survival time $T$ with the bias parameter $\eta$ for the power-law configuration model and the Erd{\H o}s-R{\' e}nyi graph for $\beta=2$. The scaling index $\alpha = 2.2$ for the SF network and the ER network is chosen to have the same average degree. The network size is $n=10^4$ vertices and the results are averaged over $10^4$ random realizations. See text for the explanation.} 
        \end{figure}

To implement this strategy in a concrete manner, we give a neighbour that is alive at time $t$ with degree $k(t)$ a share of $S$ proportional to $1/k(t)^{\eta}$ where $\eta$ is a bias parameter. The case $\eta = 0$ corresponds to the situation discussed in the previous subsection where each neighbour gets equal share. As $\eta$ is increased, more and more low degree vertices start surviving longer leading to increase in the overall survival time of the network. But this optimization cannot go on indefinitely since for very large values of $\eta$ only the lowest degree neighbours get the resources and others get nothing even if they are in need. This means that there must exist a value of $\eta$ for which the survival time gets maximized. We verify this theory by explicit numerical simulations for both scale-free and Erd{\H o}s-R{\'e}nyi cases, and the results are shown in Fig.~\ref{t_vs_eta_beta2}.

As Fig.~\ref{t_vs_eta_beta2} shows, biased strategy works effectively for SF graph but not for ER graph. This is understandable since all the vertices in ER graph have similar degrees. For SF graph however, the survival time drastically increases as more and more fraction of the surplus is diverted towards low degree vertices before the strategy becomes detrimental. In the limit of large $\eta$, only lowest degree vertices receive the benefit of surplus, and the overall survival time becomes much smaller than the unbiased strategy. 

\begin{figure}
            \centering
            \includegraphics[width=0.95\columnwidth]{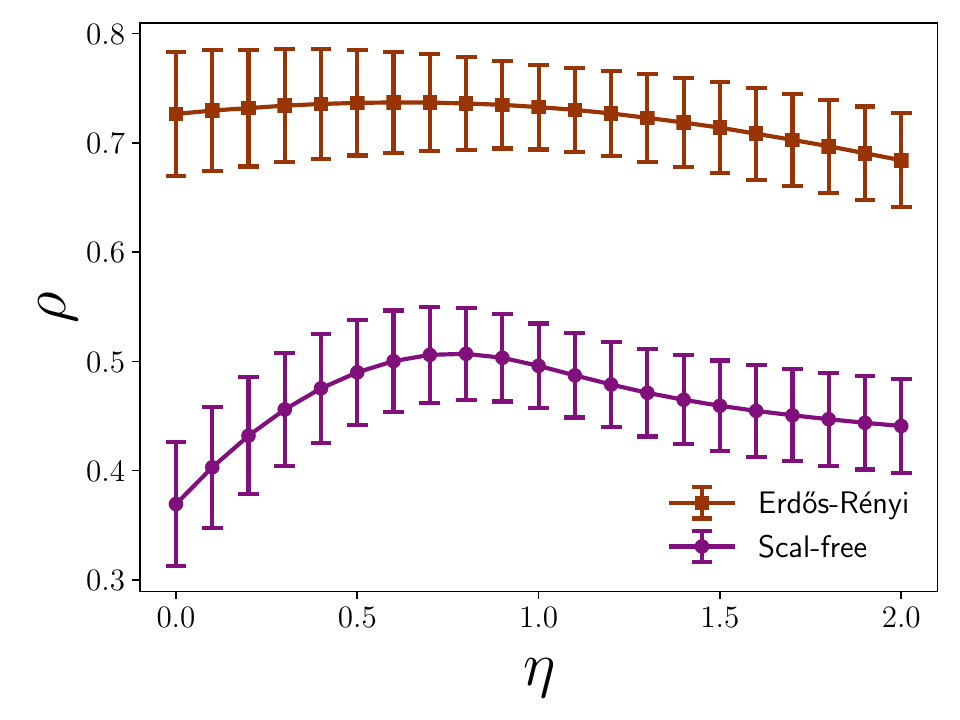}
            \caption{\label{rho_vs_eta} Efficiency index $\rho$ as a function of bias parameter $\eta$ for the two graphs. Erd{\H o}s-R{\'e}nyi graph is seen to be substantially efficient than the scale-free graph. Also, the efficiency is seen to reach the maximum for the scale-free graph in the same region where the survival time reaches the maximum confirming the presented theory (Fig.~\ref{t_vs_eta_beta2}). Error bars depict the spread of the data as measured by the standard deviation. }
        \end{figure}

We can in fact explicitly verify the theory mentioned above by looking at the survival of the individual vertices. Consider a vertex in the graph that produces amount less than $R$ at a given time. Let us call such vertex a \textit{vulnerable vertex}, since if it doesn't get resources from its neighbours, it must die. However, it may happen that a vulnerable vertex receives the required amount of resource from the neighbours, and stays safe. Let $n_{\text{vuln}}(t)$ be the number of vulnerable vertices in the network at time $t$, and let $n_{\text{saved}}(t)$ be the number of vertices saved at time $t$ among all the vulnerable ones. Note that a given vertex can become vulnerable more than once during its lifetime, and can also get saved more than once. These occurrences are counted as separate incidences while calculating $n_{\text{vuln}}$ and $n_{\text{saved}}$. Let us define the efficiency index $\rho$ of the distribution strategy as the ratio of the total number of saved and the total number of vulnerable vertices over the lifetime of the network.  
\begin{equation}
\begin{aligned}
\rho(\eta) = \frac{\sum_{t=1}^{T} n_{\text{saved}}(t)}{\sum_{t=1}^{T}n_{\text{vuln}}(t)}
\end{aligned}
\end{equation}
where $T$ is the total time for which the network survives. Clearly, $\rho \in [0, 1]$, and achieves higher value if many of the vulnerable vertices are saved on an average. It is instructive to look at the variation of $\rho$ with the bias parameter $\eta$ shown in Fig.~\ref{rho_vs_eta}. The plots clearly show that in general homogeneous degree distribution is excellent at saving the vulnerable vertices compared to the skewed degree distribution. Moreover, as $\eta$ is increased, the efficiency for the Scale-free network also increases reaching maximum around the same value of $\eta$ for which the survival time reaches maximum (Fig.~\ref{t_vs_eta_beta2}). This confirms our theory that when the resource sharing is biased, we start saving small degree vertices by diverting towards them bigger fraction of surplus that was going to the high degree vertices. This increases the overall survival time of the network. However, it also means that when $\eta$ is too large, we essentially take away all the share from high degree vertices and this decreases the survival time again. 

A ready-to-use implementation of the resource dependency model described here is freely available as a part of the package \texttt{dependency-networks} \cite{shekatkar2020dependency-networks}.

\section{\label{conclusion}Conclusion}
We have presented a simple threshold model for sharing of resources in a network where a single type of resource is produced stochastically on each vertex, and vertices share the surplus amounts with the neighbours. We studied how the network structure affects the survivability of the system with focus on the degree distribution. We did this using explicit simulations on configuration model networks with power-law and Poisson degree distributions, which are prototypes for the scale-free and homogeneous topologies respectively. In particular, we have shown that although networks with scale-free topology survive much longer owing to the presence of a few high-degree vertices than those with homogeneous topology, the later are in fact much more robust when the average survivability of the individual vertices is taken into account. This should be an important thing under consideration when building artificial networks where resource dependencies are relevant. 

The assumed finite lifetime of resources has an important consequence in the context of optimal distribution of the resources. Since the surplus resource cannot be stored indefinitely, what actually becomes important is how do we efficiently distribute the surplus so as to maximize the chances of survival. Moreover, since each vertex only knows about the topology of the neighbour, and because the sharing is local, the efficiency of the distribution cannot be increased beyond a point, and some surplus is inevitably wasted. Nevertheless, as we have shown, by biasing the distribution according to the degree of the neighbours, a substantial improvement can be achieved although no information about the production on the neighbours is known. We anticipate that these insights would be useful in a variety of real-world scenarios such as distribution of essential commodities between cities linked by roads.

The work presented here considers perhaps the simplest of the networks without edge directionality, edge weights and without other structural patterns such as degree correlations \cite{PhysRevLett.89.208701}, clustering and community structure \cite{Newman8577}. Since the real-world networks do contain many of these features, the results presented here should only be looked at as a starting point of a much broader investigation of the complex resource dependencies in real-world systems with many types of resources present. However, the approach presented here is easily generalizable to such situations by incorporating relevant structural properties in the form of general random graph models.

\begin{acknowledgments}
SMS acknowledges funding from the DST-INSPIRE Faculty Fellowship (DST/INSPIRE/04/2018/002664) by DST India. MI acknowledges the fellowship received under the same scheme to work on the project. 
\end{acknowledgments}


\begin{thebibliography}{16}
\expandafter\ifx\csname natexlab\endcsname\relax\def\natexlab#1{#1}\fi
\expandafter\ifx\csname bibnamefont\endcsname\relax
  \def\bibnamefont#1{#1}\fi
\expandafter\ifx\csname bibfnamefont\endcsname\relax
  \def\bibfnamefont#1{#1}\fi
\expandafter\ifx\csname citenamefont\endcsname\relax
  \def\citenamefont#1{#1}\fi
\expandafter\ifx\csname url\endcsname\relax
  \def\url#1{\texttt{#1}}\fi
\expandafter\ifx\csname urlprefix\endcsname\relax\def\urlprefix{URL }\fi
\providecommand{\bibinfo}[2]{#2}
\providecommand{\eprint}[2][]{\url{#2}}

\bibitem[{\citenamefont{Newman et~al.}(2006)\citenamefont{Newman, Barab{\'a}si,
  and Watts}}]{newman2006structure}
\bibinfo{author}{\bibfnamefont{M.~E.} \bibnamefont{Newman}},
  \bibinfo{author}{\bibfnamefont{A.-L.~E.} \bibnamefont{Barab{\'a}si}},
  \bibnamefont{and} \bibinfo{author}{\bibfnamefont{D.~J.} \bibnamefont{Watts}},
  \emph{\bibinfo{title}{The structure and dynamics of networks.}}
  (\bibinfo{publisher}{Princeton university press}, \bibinfo{year}{2006}).

\bibitem[{\citenamefont{Newman}(2018)}]{newman2018networks}
\bibinfo{author}{\bibfnamefont{M.}~\bibnamefont{Newman}},
  \emph{\bibinfo{title}{Networks}} (\bibinfo{publisher}{Oxford university
  press}, \bibinfo{year}{2018}).

\bibitem[{\citenamefont{Pfeffer and Salancik}(2003)}]{pfeffer2003external}
\bibinfo{author}{\bibfnamefont{J.}~\bibnamefont{Pfeffer}} \bibnamefont{and}
  \bibinfo{author}{\bibfnamefont{G.~R.} \bibnamefont{Salancik}},
  \emph{\bibinfo{title}{The external control of organizations: A resource
  dependence perspective}} (\bibinfo{publisher}{Stanford University Press},
  \bibinfo{year}{2003}).

\bibitem[{\citenamefont{Hillman et~al.}(2009)\citenamefont{Hillman, Withers,
  and Collins}}]{hillman2009resource}
\bibinfo{author}{\bibfnamefont{A.~J.} \bibnamefont{Hillman}},
  \bibinfo{author}{\bibfnamefont{M.~C.} \bibnamefont{Withers}},
  \bibnamefont{and} \bibinfo{author}{\bibfnamefont{B.~J.}
  \bibnamefont{Collins}}, \bibinfo{journal}{Journal of management}
  \textbf{\bibinfo{volume}{35}}, \bibinfo{pages}{1404} (\bibinfo{year}{2009}).

\bibitem[{\citenamefont{Davis and Cobb}(2010)}]{davis2010resource}
\bibinfo{author}{\bibfnamefont{G.~F.} \bibnamefont{Davis}} \bibnamefont{and}
  \bibinfo{author}{\bibfnamefont{J.~A.} \bibnamefont{Cobb}}, in
  \emph{\bibinfo{booktitle}{Stanford's organization theory renaissance,
  1970--2000}} (\bibinfo{publisher}{Emerald Group Publishing Limited, Bingley},
  \bibinfo{year}{2010}).

\bibitem[{\citenamefont{Buldyrev et~al.}(2010)\citenamefont{Buldyrev, Parshani,
  Paul, Stanley, and Havlin}}]{buldyrev2010catastrophic}
\bibinfo{author}{\bibfnamefont{S.~V.} \bibnamefont{Buldyrev}},
  \bibinfo{author}{\bibfnamefont{R.}~\bibnamefont{Parshani}},
  \bibinfo{author}{\bibfnamefont{G.}~\bibnamefont{Paul}},
  \bibinfo{author}{\bibfnamefont{H.~E.} \bibnamefont{Stanley}},
  \bibnamefont{and} \bibinfo{author}{\bibfnamefont{S.}~\bibnamefont{Havlin}},
  \bibinfo{journal}{Nature} \textbf{\bibinfo{volume}{464}},
  \bibinfo{pages}{1025} (\bibinfo{year}{2010}).

\bibitem[{\citenamefont{Gao et~al.}(2012)\citenamefont{Gao, Buldyrev, Stanley,
  and Havlin}}]{gao2012networks}
\bibinfo{author}{\bibfnamefont{J.}~\bibnamefont{Gao}},
  \bibinfo{author}{\bibfnamefont{S.~V.} \bibnamefont{Buldyrev}},
  \bibinfo{author}{\bibfnamefont{H.~E.} \bibnamefont{Stanley}},
  \bibnamefont{and} \bibinfo{author}{\bibfnamefont{S.}~\bibnamefont{Havlin}},
  \bibinfo{journal}{Nature physics} \textbf{\bibinfo{volume}{8}},
  \bibinfo{pages}{40} (\bibinfo{year}{2012}).

\bibitem[{\citenamefont{Brummitt et~al.}(2012)\citenamefont{Brummitt,
  D’Souza, and Leicht}}]{brummitt2012suppressing}
\bibinfo{author}{\bibfnamefont{C.~D.} \bibnamefont{Brummitt}},
  \bibinfo{author}{\bibfnamefont{R.~M.} \bibnamefont{D’Souza}},
  \bibnamefont{and} \bibinfo{author}{\bibfnamefont{E.~A.}
  \bibnamefont{Leicht}}, \bibinfo{journal}{Proceedings of the National Academy
  of Sciences} \textbf{\bibinfo{volume}{109}}, \bibinfo{pages}{E680}
  (\bibinfo{year}{2012}).

\bibitem[{\citenamefont{Zhang et~al.}(2020)\citenamefont{Zhang, Zhou, Zou,
  Tang, Xiao, and Stanley}}]{zhang2020asymmetric}
\bibinfo{author}{\bibfnamefont{H.}~\bibnamefont{Zhang}},
  \bibinfo{author}{\bibfnamefont{J.}~\bibnamefont{Zhou}},
  \bibinfo{author}{\bibfnamefont{Y.}~\bibnamefont{Zou}},
  \bibinfo{author}{\bibfnamefont{M.}~\bibnamefont{Tang}},
  \bibinfo{author}{\bibfnamefont{G.}~\bibnamefont{Xiao}}, \bibnamefont{and}
  \bibinfo{author}{\bibfnamefont{H.~E.} \bibnamefont{Stanley}},
  \bibinfo{journal}{Physical Review E} \textbf{\bibinfo{volume}{101}},
  \bibinfo{pages}{022314} (\bibinfo{year}{2020}).

\bibitem[{\citenamefont{Knuth}(1993)}]{knuth1993stanford}
\bibinfo{author}{\bibfnamefont{D.~E.} \bibnamefont{Knuth}},
  \emph{\bibinfo{title}{The Stanford GraphBase: a platform for combinatorial
  computing}} (\bibinfo{publisher}{AcM Press New York}, \bibinfo{year}{1993}).

\bibitem[{\citenamefont{Fosdick et~al.}(2018)\citenamefont{Fosdick, Larremore,
  Nishimura, and Ugander}}]{fosdick2018configuring}
\bibinfo{author}{\bibfnamefont{B.~K.} \bibnamefont{Fosdick}},
  \bibinfo{author}{\bibfnamefont{D.~B.} \bibnamefont{Larremore}},
  \bibinfo{author}{\bibfnamefont{J.}~\bibnamefont{Nishimura}},
  \bibnamefont{and} \bibinfo{author}{\bibfnamefont{J.}~\bibnamefont{Ugander}},
  \bibinfo{journal}{SIAM Review} \textbf{\bibinfo{volume}{60}},
  \bibinfo{pages}{315} (\bibinfo{year}{2018}).

\bibitem[{\citenamefont{Peixoto}(2014)}]{peixoto_graph-tool_2014}
\bibinfo{author}{\bibfnamefont{T.~P.} \bibnamefont{Peixoto}},
  \bibinfo{journal}{figshare}  (\bibinfo{year}{2014}),
  \urlprefix\url{http://figshare.com/articles/graph_tool/1164194}.

\bibitem[{\citenamefont{Spearman}(1987)}]{spearman1987proof}
\bibinfo{author}{\bibfnamefont{C.}~\bibnamefont{Spearman}},
  \bibinfo{journal}{The American journal of psychology}
  \textbf{\bibinfo{volume}{100}}, \bibinfo{pages}{441} (\bibinfo{year}{1987}).

\bibitem[{\citenamefont{Shekatkar}(2014)}]{shekatkar2020dependency-networks}
\bibinfo{author}{\bibfnamefont{S.~M.} \bibnamefont{Shekatkar}},
  \bibinfo{journal}{figshare}  (\bibinfo{year}{2014}),
  \urlprefix\url{http://figshare.com/articles/dependency-networks/13275137}.

\bibitem[{\citenamefont{Newman}(2002)}]{PhysRevLett.89.208701}
\bibinfo{author}{\bibfnamefont{M.~E.~J.} \bibnamefont{Newman}},
  \bibinfo{journal}{Phys. Rev. Lett.} \textbf{\bibinfo{volume}{89}},
  \bibinfo{pages}{208701} (\bibinfo{year}{2002}).

\bibitem[{\citenamefont{Newman}(2006)}]{Newman8577}
\bibinfo{author}{\bibfnamefont{M.~E.~J.} \bibnamefont{Newman}},
  \bibinfo{journal}{Proceedings of the National Academy of Sciences}
  \textbf{\bibinfo{volume}{103}}, \bibinfo{pages}{8577} (\bibinfo{year}{2006}),
  ISSN \bibinfo{issn}{0027-8424}.

\end{thebibliography}

\end{document}